\documentclass[final,5p,times,twocolumn]{elsarticle}

\usepackage{textcomp}
\usepackage{graphicx}
\usepackage{flafter}
\usepackage{amsmath,amssymb}
\usepackage{bm}
\usepackage[T1]{fontenc}
\usepackage[ansinew]{inputenc}
\usepackage[usenames]{color}
\newcommand{\eref}[1]{Eq.(\ref{#1})}

\newcommand{\erefs}[1]{Eqs.(\ref{#1})}

\newcommand{\qq}{\qquad}

\newcommand{\reff}[1]{(\ref{#1})}

\newcommand{\p}{\partial}

\newcommand{\pt}{\p_t}
\newcommand{\vna}{\vec\nabla}
\newcommand{\lapl}{\nabla^2}

\newcommand{\s}{\,\mathrm{s}}
\newcommand{\cm}{\,\mathrm{cm}}
\newcommand{\eV}{\,\mathrm{eV}}

\newcommand{\Mpc}{\,\mathrm{Mpc}}

\newcommand{\K}{\,\mathrm{K}}

\renewcommand{\vec}[1]{\boldsymbol{#1}}
\renewcommand{\b}{\vec{b}_1}
\newcommand{\ro}{\rho_0}
\newcommand{\ru}{\rho_1}
\newcommand{\po}{P_0}
\newcommand{\pu}{P_1}
\newcommand{\phio}{\Phi_0}
\newcommand{\phiu}{\Phi_1}
\newcommand{\bo}{\vec{B}_0}
\newcommand{\bu}{\vec{B}_1}
\newcommand{\vo}{\vec{v}_0}
\newcommand{\vu}{\vec{v}_1}
\newcommand{\hbo}{\hat{\vec{B}}_0}
\newcommand{\hq}{\hat{\vec{q}}}

\newcommand{\para}{{\scriptscriptstyle{\parallel}}}
\newcommand{\perpt}{{\scriptscriptstyle{\perp}}}
\newcommand{\vup}{v_{1}^\para}
\newcommand{\vuo}{\vec{v}_{1}^\perpt}

\newcommand{\bob}{\bar{\vec{B}}_0}
\newcommand{\kv}{\vec{k}}
\newcommand{\rv}{\vec{r}}
\newcommand{\bb}{\bar{b}}
\newcommand{\vb}{\bar{v}}
\newcommand{\rob}{\bar{\rho}}

\newcommand{\lu}{\Lambda_1}
\newcommand{\ld}{\Lambda_2}
\newcommand{\ie}{\emph{i.e.}, }
\newcommand{\zrec}{z_{rec}}
\newcommand{\amb}{\mathrm{amb}}
\newcommand{\Ramb}{R_\amb}
\newcommand{\Lamb}{L_\amb}

\newcommand{\tamb}{\tau_\amb}
\journal{Physics Letter B}

\begin{document}

\begin{frontmatter}



\title{Gravitational instability of the primordial plasma: anisotropic evolution of structure seeds}


\author[Sapienza,Ox,Mib,Fe]{Massimiliano Lattanzi} \ead{lattanzi@fe.infn.it}
\author[Sapienza]{Nakia Carlevaro}
\author[ENEA,Sapienza,INFN]{Giovanni Montani}

\address[Sapienza]{Dipartimento di Fisica, ``Sapienza'' Universit\`a di Roma, P.le A. Moro 5 (00185), Roma, Italy.} 
\address[Ox]{Physics Department, University of Oxford, OX1 3RH Oxford, United Kingdom.}
\address[Mib]{Dipartimento di Fisica G. Occhialini, Universit\`a Milano-Bicocca and
INFN, Sezione di Milano-Bicocca, Piazza della Scienza 3, I-20126 Milano, Italy.}
\address[Fe]{Dipartimento di Fisica, Universit\`a di Ferrara and INFN, sezione di Ferrara, 
Polo Scientifico e Tecnologico - Edificio C Via Saragat, 1, I-44122 Ferrara Italy}
\address[ENEA]{ENEA, C.R. Frascati (Rome), UTFUS-MAG, Italy.}
\address[INFN]{INFN, Sezione Roma1, Italy}

\begin{abstract}

We study how the presence of a background magnetic field, of intensity compatible with current observation constraints, affects the linear evolution of cosmological density perturbations at scales below the Hubble radius. The magnetic field provides an additional pressure that can prevent the growth of a given perturbation; however, the magnetic pressure is confined only to the plane orthogonal the field. As a result, the ``Jeans length'' of the system not only depends on the wavelength of the fluctuation but also on its direction, and the perturbative evolution is anisotropic. We derive this result analytically and back it up with direct numerical integration of the relevant ideal magnetohydrodynamics equations during the matter-dominated era. Before recombination, the kinetic pressure dominates and the perturbations evolve in the standard way, whereas after that time magnetic pressure dominates and we observe the anisotropic evolution. We quantify this effect by estimating the eccentricity $\epsilon$ of a Gaussian perturbation in the coordinate space that was spherically symmetric at recombination. For a perturbations at the sub-galactic scale, we find that $\epsilon = 0.7$ at $z=10$ taking the background magnetic field of order $10^{-9}$ gauss.

\end{abstract}


\end{frontmatter}

% \linenumbers

\section{General Remarks}
 
Our theoretical knowledge of the Universe is based on the Standard Cosmological Model, that provides a convenient framework to satisfactorily explain the majority of cosmological observations, like the anisotropy pattern of the cosmic microwave background (CMB) \cite{Komatsu:2010fb,Larson:2010gs,Dunkley:2010ge}, the large-scale structure of the Universe \cite{Tegmark:2004,Cole:2005sx,Tegmark:2006az}, the Hubble diagram of distant type Ia supernovae \cite{Riess:1998cb,Perlmutter:1998np,Frieman:2008sn} and the abundances of light elements \cite{Iocco:2008va}.
 
The Standard Cosmological Model relies on the assumption that the Universe, at least at large scales, is highly homogeneous and isotropic, and its geometry is thus described by the Robertson-Walker metric. In fact, the distribution of luminous red galaxies shows that the present Universe is homogeneous on scales greater than $\sim 100$ Mpc \cite{Hogg:2004vw}, while the isotropy of the CMB itself (which has a black-body distribution at $T=2.73\,$K with temperature fluctuations of order $10^{-5}$ or less) is an indication of the isotropy of the Universe as a whole and a strong evidence for homogeneity at the time of hydrogen recombination (nearly 400.000 years after the Big Bang, corresponding to a cosmological redshift $z_\mathrm{rec}=1100$). On the other hand, below the ``homogeneity scale'' of 100 Mpc, the distribution of matter is definitely inhomogeneous. Such a dichotomy between the smoothness in the matter-energy distribution at $z=\zrec$ and the clumpiness of the recent Universe (for $z\lesssim1$) below a certain scale is explained by the mechanism of gravitational instability: the structures we observe today have been formed through the growth of tiny density perturbation seeds that, accordingly to the currently accepted model, were created in the early Universe during a phase of inflationary expansion.

The presence of a large scale (\ie coherent over a Hubble length), strong magnetic field is forbidden by the observed isotropy, as it would naturally single out a preferred spatial direction. However, a background and uniform magnetic field could be present at cosmological scales provided that its intensity is small enough. In particular, upper limits on the present field intensity of order $\sim 10^{-9}$ G have been derived from observations of the CMB temperature anisotropies \cite{BFS97,Paoletti:2010rx} and of its temperature-polarization correlation \cite{Scannapieco:1997mt,Komatsu:2010fb}\footnote{It has been argued (see, e.g., \cite{Adamek:2011pr}) that the limits obtained from the  CMB temperature data can be significantly relaxed in the presence of free-streaming neutrinos. However this does not affect the limits from the polarization.}. Smaller scale fields, on the other hand, could be as strong as $10^{-6}$ G.

The effects of large-scale magnetic fields on the evolution of cosmological structures have been studied extensively in the literature (for a complete review, see Ref. \cite{BMT07} and references therein), where both Newtonian and general relativistic treatments, the latter often using covariant and gauge-invariant techniques, are present. It is known that, among others, magnetic fields slow down (and possibly prevent) the growth of perturbations and can produce vorticities and shape distortions in the density field \cite{VTP05,Tsagas:1999ft}. In this Letter, our goal is to revisit the issue of the existence of a ``magnetic Jeans length'' and of its dependence on the direction along which the perturbation propagates, other than to give a realistic estimate of its value in a matter-dominated Universe. The presence of a magnetic Jeans length has been discussed in several papers \cite{BMT07,VTP05,Tsagas:1999ft,RR71, Kim:1994zh} both in a Newtonian and in general relativistic framework (for an analysis of the standard Jeans mechanism in the presence of dissipative effect see also \cite{CQG,IJMPD,MPLA}), but some of the analyses failed to recognize its angular dependence. Here, we present a neat derivation of the relevant instability scales bases on the equations of magnetohydrodynamics (MHD) on an expanding Universe. We discuss a simple generalization of the magnetic Jeans length suited for two-fluid systems, and clarify some misunderstandings that are present in the literature. We also show the results of the numerical integration of the coupled MHD and Poisson equations. Finally, we numerically study the distortion introduced in the density field by the anisotropy in the critical length.

The Letter is organized as follows. In Sec. 2, we characterize the Universe as a plasma. In Sec. 3, we introduce the basic equations, and in Sec. 4 we carry out the linearization procedure. We derive the existence of the magnetic Jeans length from analytical considerations in Sec. 5, while in Sec. 6 we show some numerical results. Finally, we draw our conclusions in Sec. 7.

\section{The plasma features of the pre- and post-recombination Universe}

In this Section, we aim at characterizing the plasma features of the cosmological fluid. Between the time of $e^+ e^-$ annihilation (\ie for a temperature\footnote{All throughout the paper, we use natural units with $c=\hbar=k_B=1$.} $T\simeq m_e$, corresponding to a redshift $z\sim 10^9$) and the present ($z=0$), the matter-energy content of the Universe is provided by electrons, protons and neutrons (the three species being collectively referred to as baryons in the cosmological jargon), photons, neutrinos, and two elusive components dubbed dark matter and dark energy, that presently account for more than 99\% of the total energy budget of the Universe. However, dark energy has been subdominant for most of the past history of the Universe and can be safely neglected at redshifts $z>1$. Dark matter and neutrinos interact only gravitationally with the other components and can be neglected as long as the plasma properties of the fluid are concerned.

The cosmological baryon-to-photon ratio is extremely small and equal to $n_b/n_\gamma\simeq6.1\times 10^{-10}$, where $n_\gamma$ and $n_b$ are the photon and baryon number densities, respectively. Both $n_b$ and $n_\gamma$ scale with redshift as $(1+z)^3$; their present values are $n_{b}(z=0) \simeq2.5\times 10^{-7} \cm^{-3}$ and $n_\gamma(z=0)\simeq 410 \cm^{-3}$. Most of the baryons in the Universe are in the form of $^1\mathrm H$ nuclei (i.e., isolated protons) so that, for simplicity, in the following we assume $n_b=n_p$ ($n_p$ being the proton number density). Furthermore, $n_p$ is also equal to the electron number density $n_e=n_p$, because of the charge neutrality of the Universe. 

We study separately the properties of the cosmological fluid before and after the time of hydrogen recombination occurring at $z=\zrec=1100$ ($T\simeq 0.25\eV$). For $z>\zrec$, the protons and electrons are free and thus one deals with a fully ionized plasma. In this regime, photons and baryons are tightly coupled due to Thomson scattering, and share a common temperature $T(z)=T_\gamma(z)=T_\gamma^0 (1+z)$, where the present photon temperature $T_\gamma^0=2.73\,\K$.  After recombination ($z<\zrec$), most of the electrons and protons exist in the form of neutral hydrogen atoms, and only a small residual ionized fraction $x_e=2.5\times 10^{-4}$ survives, making the fluid a weakly ionized plasma. In this regime, the photon temperature  still scales as $(1+z)$, while that of baryons evolves in the same way only until $z=100$, due to residual scatterings that keep them in thermal equilibrium with photons; after that time, their temperature decreases faster, as $(1+z)^{2}$. The baryonic fluid remains neutral until the time of reionization, when the UV radiation produced by the first stars ionizes again the hydrogen present in the cosmological medium. This is likely to have happened around $z\simeq 10$, however the precise details of the reionization history are still largely unknown, and for this reason we limit our analysis to redshifts $z>10$.

In the following, we will assume the presence of a background homogeneous magnetic field $\vec{B}(z)$, whose contribution to the total energy density of the Universe can be considered negligible. We recall that the field intensity $B(z)$ scales as $(1+z)^2$ and, unless otherwise stated, we take its value at the present time to be $B(z=0)=10^{-9}$ G.

\subsection{The pre-recombination Universe}
A fundamental quantity characterizing a plasma is the Debye length $\lambda_D$, namely the length over which electrons screen out electric fields in a plasma. It defines the length scale over which a system can consistently considered to be a plasma. The Debye length of a hydrogen plasma at temperature $T$ is 
\begin{equation}
\lambda_D=\sqrt{\frac{T}{4\pi n_e e^2}}\simeq(6.9\cm)\sqrt{\frac{T/\mathrm{K}}{n_e/\mathrm{cm}^{-3}}}\;,
\end{equation}
where $e$ is the proton charge. Using $T=T_\gamma$ and the values given above, one gets
\begin{equation}
\lambda_D(z)=\frac{2.3\times 10^4 \cm}{(1+z)}\;.
\label{eq:ldz}
\end{equation} 
The redshift dependence of $\lambda_D$ implies that the comoving Debye length $\bar\lambda_D\equiv\lambda_D (1+z)$ is constant during the cosmological evolution and equal to $\bar\lambda_D\simeq 2 \times 10^4\cm$. 

Plasma effects can be important in a system when its physical dimension $L$ is much larger than the Debye length. For the Universe, the relevant length is the Hubble radius $L=l_H\equiv H^{-1}$, %c
where $H$ is the Hubble parameter. This length represents the maximum scale at which microphysical processes can operate in order to establish the thermodynamical equilibrium. Today, $l_H\simeq 10^{28}\cm$; during the matter-dominated era $l_H\propto (1+z)^{-3/2}$, while in the radiation-dominated era $l_H\propto(1+z)^{-2}$. From the analysis of both these scales, it is evident that $l_H\gg\lambda_D$ turns out in the period considered here. Moreover, under the same hypotheses, the baryonic matter $M_D$ within a Debye sphere is also constant and given by
\begin{equation}
M_D = \tfrac{4}{3}\,\pi m_p n_b\lambda_D^3\simeq  10^{-50} M_{\odot}\;,
\end{equation}
where $m_p$ is the proton mass. This ``Debye mass'' clearly results to be much smaller than any other of cosmological interest and we can conclude that the cosmological fluid can be considered as neutral at all relevant scales.

Another meaningful index is the so-called plasma parameter $N_D$, \ie the number of particles within a Debye sphere:
\begin{equation}\label{eq:nd}
N_D= \tfrac{4}{3}\,\pi n_b\lambda_D^3 \;.
\end{equation}
The dependence  of $\lambda_D$ and $n_p$ on the redshift $z$ implies that also $N_D$ is a constant. In particular, since $N_D\simeq 10^7\gg1$, the cosmological fluid results to be a weakly coupled plasma.

In order to provide a complete characterization of the cosmological plasma, we now turn our attention to the plasma dissipative properties, starting from the plasma resistivity $\eta$. For an electron-proton plasma, this is given by $\eta=m_e \nu_{ei}\;/\;n_e e^2$, where $m_e$ is the electron mass and $\nu_{ei}$ is the electron-ion collision frequency. For the case under consideration, $\nu_{ei}$ is well approximated by the electron-electron collision frequency $\nu_{ee}$ \cite{plasmaf}, \ie
\begin{equation}
\nu_{ei}\simeq \nu_{ee} \simeq(2.91\times 10^{-6} \,\mathrm{s}^{-1}) \left(\frac{n_e}{\cm^{-3}}\right)\left(\frac{T}{\eV}\right)^{-3/2}\ln \Lambda_C\;,
\label{eq:nu_ei}
\end{equation}
where $\ln\Lambda_C$ is the Coulomb logarithm, introduced to quantify the effects that small-angle-diffusion collisions have in the Coulomb scattering. A simply estimate of $\Lambda_C$ in a plasma is given by $\Lambda_C\simeq12\pi N_D$, so that for the cosmological fluid, the Coulomb logarithm is $\simeq 20$. Substituting \eref{eq:nu_ei} into the expression for the resistivity given above, we get
\begin{equation}\label{etaz}
\eta(z)\simeq 1.6\times \left(\frac{1+z}{1+\zrec}\right)^{-3/2} \Omega \,\cm \;.
\end{equation}
Close to recombination, the cosmological plasma has an electric resistivity equal to $\eta(\zrec)\simeq1.6 \,\Omega \cm$, \ie a conductivity $\simeq 0.6$ siemens $\cm^{-1}$, a value typical of a semiconductor [in Gaussian units, $\eta(\zrec) = 1.8\times 10^{-12} \s $].

Let us now turn our attention to the viscous properties of the plasma. The shear viscosity coefficient of matter strongly coupled with radiation can be expressed as \cite{G&C,W71}
\begin{equation}
\eta_v=\frac{4}{15}\;a_{SB} T^4 \tau\;,
\end{equation}
where $a_{SB}\simeq 5.7\times10^{-8}\,\mathrm{W}\,\mathrm{K}^{-4}\mathrm{m}^{-2}$ is the Stefan-Boltzmann constant, while $\tau$ denotes the mean collision time between particles and can be estimated as $\tau\simeq(n_\gamma\sigma_T v)^{-1}$ (here, $v\simeq c$ and we have introduced $\sigma_T\simeq6.6\times10^{-29}\mathrm{m}^2$ as the cross section for Thomson scattering). For a photon gas at equilibrium at temperature $T$, $n_\gamma\simeq2.0\times10^7(T/\mathrm{K})^3\mathrm{m}^{-3}$ and then
\begin{equation}
\eta_v(T) \simeq 1.2 \times 10^{-4}\left(\frac{T}{\mathrm{K}} \right) \frac{\mathrm{kg}}{\mathrm{m\,s}}\;.
\end{equation}

The resistivity and viscosity coefficients enter the MHD equations through the following diffusion coefficients
$\bar\eta\equiv\eta/4\pi$ %c^2 
and $\bar\eta_v\equiv\eta_v/\rho$, where $\rho$ is the density of the fluid. Taking $\rho=\rho_b=m_pn_b \simeq 4.2\times 10^{-28}(1+z)^3$ kg m$^{-3}$ and using  $T=T_\gamma^0(1+z)$, we get
\begin{align}
&\bar\eta_v \simeq (6.4 \times 10^{17}\,\mathrm{m^2\,s^{-1}}) \left(\frac{1+z}{1+z_{rec}}\right)^{-2}\;, \\
&\bar\eta     \simeq (1.3 \times 10^4\,\mathrm{m^2\,s^{-1}})\; \left(\frac{1+z}{1+z_{rec}}\right)^{-3/2}\;.
\end{align}
The relative magnitude of the viscous and magnetic diffusion rates can be parameterized through the magnetic Prandtl number $Pr_m \equiv \bar\eta_v/\bar \eta$. Using the expression above for $\bar \eta_v$ and $\bar\eta$, we obtain
\begin{equation}
Pr_m \simeq 5.0\times 10^{13} \left(\frac{1+z}{1+z_{rec}}\right)^{-1/2}\;,
\end{equation}
so that $Pr_m \gg 1$, \ie viscous diffusion is more important than resistive diffusion, at recombination and indeed always at the redshifts under consideration.

Having discussed the relative importance of viscosity- and resistivity-driven dissipative effects, we now analyze at which scales these effects are relevant. In a magnetized plasma, a useful parameter is the Lundquist number $S \equiv L v_A/\eta$, where $L$ is a typical length scale and $v_A=(B/4\pi\rho)^{1/2}$ is the Alfv\'en velocity. The Lundquist number is basically the ratio between the resistive diffusion timescale $\tau_r= L^2/\bar\eta$ and the Alfv\'en crossing timescale $\tau_A = L/v_A$. Assuming $B(z=0)=10^{-9}\,\mathrm{G}$ and $\rho=\rho_b$, we obtain $v_A\simeq1.2\times10^{5}\,\mathrm{m/s}\,[(1+z)/(1+z_{rec})]^{1/2}$ and thus
\begin{equation}
S = \frac{\tau_r}{\tau_A}=2.8\times 10^{23} \left(\frac{L}{\Mpc}\right)\left(\frac{1+z}{1+z_{rec}}\right)^2\,.
\end{equation}
We also compare $\tau_A$ with the viscous diffusion timescale defined as $\tau_v = L^2/\bar\eta_v$. The ratio $S_v\equiv\tau_v/\tau_A$ can be thought as a viscous analogous to the Lundquist number: 
\begin{equation}
S_v = \frac{\tau_v}{\tau_A}=5.7\times 10^{9} \left(\frac{L}{\Mpc}\right)\left(\frac{1+z}{1+z_{rec}}\right)^2\;.
\end{equation}
The baryonic mass (at the average background density) contained in a sphere of radius equal to the length where $S\sim1$ can be calculated to be $\sim 10^{-51} M_{\odot} [(1+z)/(1+z_{rec})]^{-3}$, while the analogous quantity for $S_v$ is $\sim 10^{-10} M_\odot [(1+z)/(1+z_{rec})]^{-9/2}$. Both mass scales are well below the values of cosmological relevance at the redshifts of interest.

The discussion above shows how the following hierarchy among the relevant time scales holds for all mass range and redshifts of interest:
\begin{equation}
\tau_A \ll \tau_v \ll \tau_r \;,
\label{eq:hierdiss}
\end{equation}
meaning that viscosity always dominates over resistivity, and that both dissipative effects can indeed be neglected when studying the propagation of Alfv\'en waves.

\subsection{The post-recombination Universe \label{ssec:post_uni}}

After the time of recombination $z_{\textrm{rec}}=1100$, the cosmological plasma exists in a weakly ionized state, the neutral and ionized components having densities $\rho_n\simeq \rho_b$ and $\rho_i=x_e\rho_b\ll \rho_n$, respectively. We take the residual ionization fraction $x_e$ constant and equal to $2.5\times 10^{-4}$ in the range $\zrec>z>10$. The results of the previous Subsection can be generalized to show that the hierarchy \reff{eq:hierdiss} holds also in this regime.

In spite of the small value of the ionization fraction, the magnetic field could still affect the dynamics of the whole system in view of the interactions between neutral and charged particles. In particular, the magnetic forces acting on the charged particles can be communicated to the neutrals through collisions. However, if the coupling is not tight enough, the neutrals feel the magnetic field but drift with respect to the ions in a process termed \emph{ambipolar diffusion}. Its relevance at a given length scale $L$ is quantified by the ambipolar Reynolds number $\Ramb$ \cite{MS56,Sh83,BJ04,LM06}
\begin{equation}
\Ramb(L) \equiv \frac{v \,\gamma_\mathrm{in}\,x_e\,\rho_n}{v_A^2} L=\frac{L}{\Lamb}\; ,
\label{eq:Ramb}
\end{equation}
where $\gamma_\mathrm{in}=1.9\times 10^{-9} \mathrm{cm}^3\,\mathrm{s}^{-1}$ \cite{DR83} is the ion-neutral drag coefficient due to collisions between the two species, $v_A^2=B^2/4\pi \rho_n$ is the Alfv\'en velocity in the tightly coupled limit, $v$ is the characteristic velocity of the fluid, and $\Lamb$ is the ambipolar length, i.e., the scale where $\Ramb=1$. The ambipolar Reynolds number is just the ratio between the ambipolar diffusion timescale $\tamb=L^2/(\tau_\mathrm{ni} v_A^2)=\tau_A^2/\tau_\mathrm{ni}$ (where $\tau_\mathrm{ni}=(\gamma \rho_i)^{-1}$ is the neutral collision timescale) and some characteristic timescale $\tau=L/v$.

If $\Ramb\ll 1$, the neutrals are uncoupled from the plasma. On the contrary, when $\Ramb\gtrsim 1$ the dynamics of the two components can be described through ordinary single-fluid MHD with an additional dissipative term \cite{BJ04}. This term becomes progressively less important as $\Ramb$ grows and can be neglected in the limit $\Ramb\gg 1$, or $L\gg \Lamb$.

Assuming that the evolution of the fluid is driven by Alfv\'enic phenomena, i.e., $v \sim v_A$, the ambipolar length is given by
\begin{equation}
\Lamb= (1.2\,\mathrm{Mpc})\,(1+z)^{-5/2}\;,
\label{eq:lamb}
\end{equation}
and $\Ramb=\tamb/\tau_A$. Thus, $\Ramb \gg 1$ if and only if the tight-coupling condition $\tau_\mathrm{ni}\ll\tau=\tau_A$ is satisfied. 

\begin{figure}[t]
\centering
\includegraphics[width=0.9\hsize]{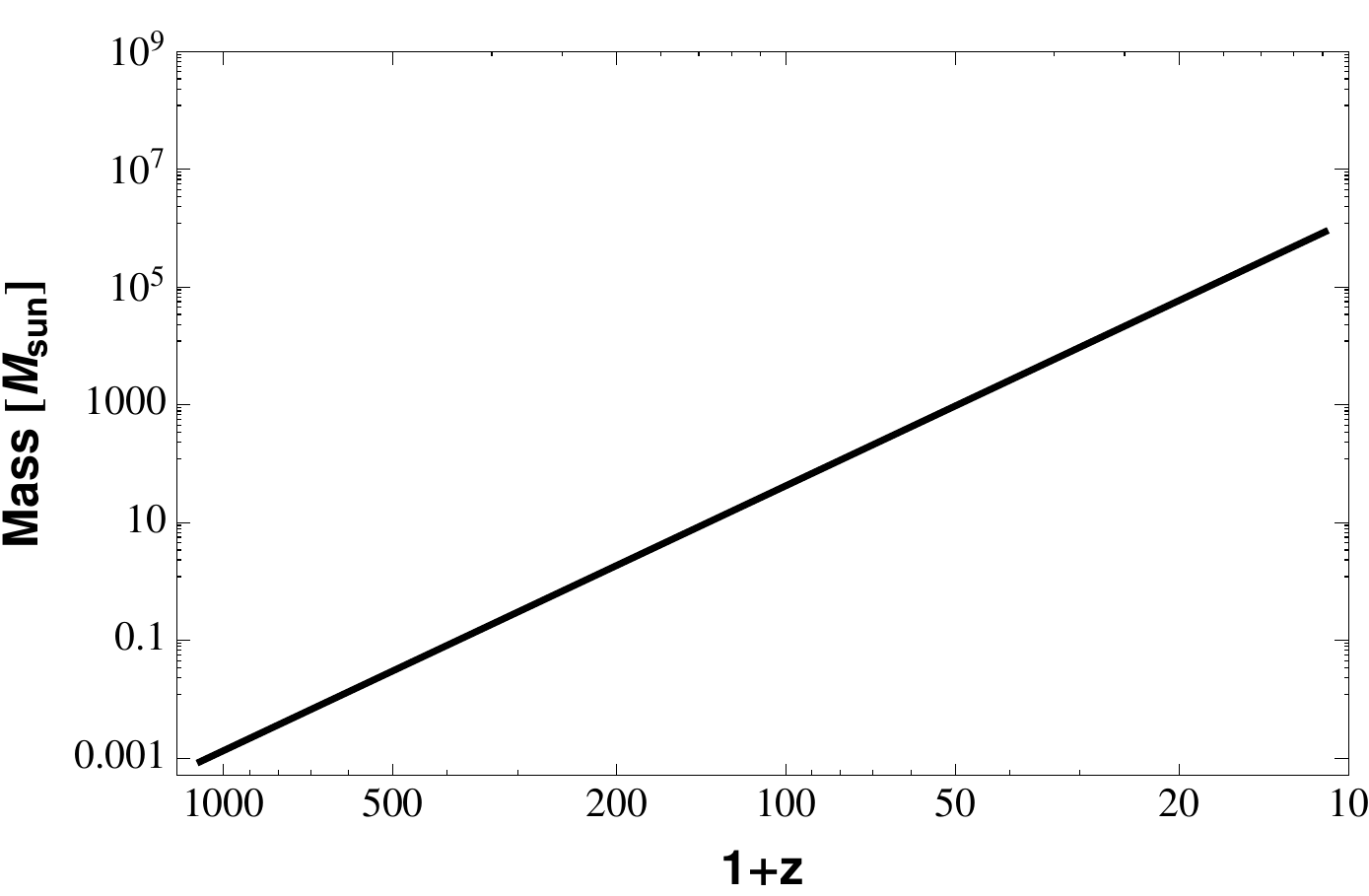}
\caption{Mass contained (at the background baryon density) within the scale $\Lamb$ defined in \eref{eq:lamb} as a function of redshift $z$. Above line, the condition $\tamb>\tau_A$ holds.\label{fig:lamb}}
\end{figure}
It is straightforward to check that for $1100>z>10$, it is always $\Ramb >1$ at scales larger than a few tens of comoving kiloparsecs, meaning that the following hierarchy holds:
\begin{equation}
\tau_\mathrm{ni} \ll \tau_A \ll \tamb \,,
\end{equation}
so that the ions and neutrals are tightly coupled and ambipolar diffusion can be safely neglected. In order to better illustrate this point, in Figure \ref{fig:lamb} we plot the mass contained in a sphere of radius $\Lamb$ at the background baryon density as a function of redshift. It is evident that, in the redshift range considered, $\Ramb\gg 1$ for all scales $M\gg 10^6 M_\odot$. 

In this regime, we can therefore neglect the dissipative term mentioned above and use single-fluid ideal MHD. We conclude that ambipolar diffusion does not affect the dynamics of the cosmological plasma after recombination.

\section{Basic equations}
In this Section, we derive the basic equations describing the linear evolution of instabilities in the cosmological fluid, modeled as a magnetized plasma. In this respect, we underline that the full investigation of the perturbative dynamics of the Universe would require a general-relativistic treatment, in order to correlate the matter and geometrical fluctuations \footnote{For a complete derivation of the Vlasov theory on curved spacetime, see Ref. \cite{DF10}.}. However, as long as one is interested in scales much smaller than the Hubble radius, \ie $L\ll H^{-1}$, %c
a Newtonian treatment provides a consistent description of the dynamics. Nonetheless, in this scenario the expansion of the Universe can be accounted as the bulk background motion of the fluid \cite{G&C,E&U,P&C}.

The starting point of our treatment is the Eulerian set of equations governing the fluid motion, on which one can develop a perturbative theory by adding small fluctuations to the unperturbed cosmological background solution. The zeroth-order dynamics is derived by considering a flat homogeneous and isotropic Universe whose energy density is dominated by non-relativistic matter, and correctly describes the expansion of the Universe.
We assume that a background magnetic field is present, whose contribution to the total energy density of the Universe can be considered negligible.

Let us now start by briefly recalling the basic equations of non-relativistic, ideal and single fluid MHD, which govern the plasma motion. The mass conservation and the Newtonian gravitational field are described by the continuity and Poisson equations, the single-fluid dynamics is described by the Euler equation in presence of a magnetic field $\vec{B}$ and, finally, the electromagnetic interaction can be summarized by the frozen-in and the Gauss laws. Such equations read
\begin{subequations}\label{initial-system}
\begin{align}	
\pt\rho+\vna\cdot\rho\vec{v}=0&\;,\label{continuity}\\
\lapl\Phi-4\pi G\rho=0&\;,\label{poisson}\\
\rho\pt\vec{v}+\rho(\vec{v}\cdot\vna)\vec{v}+\vna P+\qq\qq\qq\qq&\nonumber\\
+\rho\vna\Phi-(\vna\times\vec{B})\times\vec{B}/4\pi=0&\;,\label{eq:Euler}\\
\pt\vec{B}-\vna\times(\vec{v}\times\vec{B})=0&\;,\label{eq:base2}\\
\vna\cdot\vec{B}=0&\;,\label{eq:Gauss}	
\end{align}
\end{subequations}
respectively, where $\rho$ is the mass density, $\vec{v}$ is the velocity field, $\Phi$ is the gravitational potential and $G$ is Newton constant. This system constitutes the base of our perturbative approach. 

To derive the zeroth-order dynamics, we assume the usual Robertson-Walker metric, \ie $ds^2=dt^2 - a^2(t)\,d\ell^2$, where $a=a(t)$ represents the cosmological scale factor, and a perfect fluid energy-momentum tensor as the matter source of the gravitational field, \ie ${T_{\mu}}^{\nu}=\mbox{diag}\,[\,\ro,\,-\po,\,-\po,\,-\po\,]$, with $\ro=\ro(t)$. In this scheme, the behavior of the mass density with time is obtained from the energy-momentum conservation law $T_{0;\,\nu}^{\,\nu}=0$ and from the Friedmann equation, \ie
\begin{align}
&\dot{\rho}_0+3H(\ro+\po)=0\;,\quad\quad \label{EMT}\\
&\dot{a}^2+\mathcal{K}-\tfrac{8}{3}\pi G\ro a^2=0\;, \label{Friedmann}
\end{align}
respectively (the dot ($\dot{\;\;}$) denotes the total derivative with respect to synchronous time). Here $H=\dot{a}/a$ is the Hubble parameter and $\mathcal{K}=const.$ is the curvature factor. 

Setting the matter-dominated Universe equation of state (EoS) $\po\sim0$ ($\po\ll\ro$) in \eref{EMT}, the zeroth-order solution of the system \reff{initial-system} turns out to be
\begin{align}\label{zeroth-order-sol}
\ro=\frac{\rob}{a^3}\;,\;\;\;
\vo=H\rv\;,\;\;
\bo=\frac{\bob}{a^2}\;,\;\;\;
\vna\phio=\tfrac{4}{3}\pi G \ro \rv\;,
\end{align}
where $\rob$ and $\bob$ are dimensional constants, $\rv$ ($r=\mid\!\!\rv\!\mid$) denotes the radial coordinate vector and, of course, $a(t)$ satisfies \eref{Friedmann}. We observe how this non-stationary solution characterizing the background dynamics is not affected by the so-called ``Jeans swindle'' proper of the static solution \cite{G&C,E&U,P&C}.

To obtain now the explicit time dependence of the unperturbed quantities involved in the model, we restrict the analysis to the flat case, \ie $\mathcal{K}=0$. From the Friedmann equation \reff{Friedmann} and using the solution for $\ro$, one readily obtains
\begin{subequations}\label{time-parameters}
\begin{align}
a&=\Big(6\pi G\rob\Big)^{1/3} t^{2/3}\;, \\
\ro&=\frac{1}{6\pi G t^{2}}\; .\label{rho_t}
\end{align}
Finally, we recall that the adiabatic sound speed is defined by $v_s=\sqrt{\partial P/\partial \rho}$. For a general specific heat ratio $\gamma$, we assume that the pressure varies as $P=K \rho^{\gamma}$, so that the speed of sound is given by
\begin{equation}\label{time-vs}
v_s^{2}=\gamma\,K \,\ro^{\gamma-1}=\frac{\gamma\,K}{(6\pi\, G)^{\gamma-1}}\,t^{-2\gamma+2}\;.
\end{equation}
\end{subequations}

\section{Perturbation scheme}
In order to analyze the implications that the physics of an ideal magnetized plasma can have on the structure formation, we will follow the standard perturbation approach. In this respect, we consider small perturbations around the zeroth-order cosmological solution derived above, \ie we write $\rho= \ro+\ru$ (with $\ru\ll\ro$) and similarly for the other quantities $P$, $\vec v$, $\Phi$ and $\vec B$. Substituting the perturbed quantities in \erefs{initial-system} and keeping only terms up to first order, one gets
\begin{subequations}\label{perturbation-system}
\begin{align}
\pt\ru+3H\ru+H(\vec{r}\cdot\vna)\ru+\ro\vna\cdot\vu&=0\\
\lapl\phiu-4\pi G\ru&=0\\
\pt\vu+ H\vu+
H(\vec{r}\cdot\vna)\vu+v_s^2\vna\ru/\ro+\;\;\nonumber\\
+\vna\phiu-(\vna\times\bu)\times\bo/(4\pi\ro)&=0\\
\pt\bu+2H\bu+H(\vec{r}\cdot\vna)\bu+\quad\qq\qq\quad\nonumber\\
+\bo(\vna\cdot\vu)-(\bo\cdot\vna)\vu&=0\\
\vna\cdot\bu &=0
\end{align} 
\end{subequations}
where, as already discussed, the pressure and density perturbations have been related through the adiabatic sound speed, \ie $\pu=v^2_s\ru$. We are assuming that $B_0^2/4\pi \rho_0=v_A^{2}\ll 1$, where $B_0=|\bo|$, in order to preserve the isotropy of the background flow.

In the following, we replace $\bu$ with the dimensionless magnetic fluctuation $\b\equiv\bu/B_0$. Moreover, the analysis of the system above can be simplified by Fourier-transforming the spatial dependence of the involved quantities, \ie using perturbations in the form of plane waves, taking
\begin{align}
\phi_1(\rv,t)=\tilde{\phi}_1(t)e^{i \kv\cdot\rv}\;,
\end{align}
with $\phi_1 = \left\{\ru,\,\vu,\,\phiu,\,\bu \right\}$ and $\kv$ is the physical wavenumber scaling as $1/a(t)$. It is convenient to consider also the comoving wavenumber $\vec q = a\kv$, that stays constant during the expansion. The evolution for a given harmonic can be obtained by the equations in real space with the substitutions $\phi_1\to\tilde\phi_1$, $\vna \to i\kv$ and $\pt \to \pt - i H (\kv\cdot\rv)$. In the following, for the sake of simplicity, we will drop the tilde over the Fourier transformed variables. Then, the system \reff{perturbation-system} reduces to (hats denote unit vectors): 
\begin{subequations}
\begin{align}
\dot{\rho}_1+3H\ru+i\ro(\kv\cdot\vu)&=0\;,\\[0.2cm]
\dot{\vec v}_1+H\vu+i\left[\frac{v_s^2}{\ro}-\frac{4\pi G }{k^2}\right]\ru\kv%+\;\;%&\nonumber\\
+iv_A^2\,\hbo\times(\kv\times\b)&=0\;,\\[0.2cm]
\dot{\vec b}_1+
i\hbo(\kv\cdot\vu)-i(\hbo\cdot\kv)\vu&=0\;,
\end{align}
\end{subequations}
where we have already eliminated $\phiu$ by means of the Poisson equation in $k-$space, \ie $k^2\phiu=-4\pi G\ru$. It is understood that the constraint $\kv\cdot\b =0$ always hold.

Decomposing now $\vu$ in its components $\vup$ and $\vuo$ parallel and orthogonal to the direction of $\vec q$ respectively, \ie $\vu=\vup\,\hq+\vuo$ (where $\vuo\cdot\hq=0$), and introducing the following scalar variables:
\begin{subequations}
\begin{align}
\delta&\equiv\ru/\ro\;,\phantom{(\b\cdot\hbo)} \theta\equiv i (\kv\cdot\vu) = i k \vup\;,\\
\bb&\equiv\,(\b\cdot\hbo)\;,\phantom{\ru/\ro}\vb\equiv i\,k\, (\vuo\cdot\hbo)\;,
\end{align}
\end{subequations}
we finally get a further simplified system:
\begin{subequations}\label{master-system}
\begin{align}
\dot{\delta}+\theta=0\;,\\
\dot{\theta}+2H\theta-\omega_0^2\delta-\omega_A^2\bb=0\;,\\
\dot{\bb}+
(1-\mu^2)\theta - \mu \vb =0\;,\\
\dot{\vb}+2H\vb+\mu\omega_A^2\bb=0\;,
\end{align}
\end{subequations}
where we have defined
\begin{align}
\mu\equiv\hbo\cdot\hq\;,\qquad
\omega_A^{2}\equiv v_A^2 k^2\;,\qquad
\omega_0^{2}\equiv v_s^{2}k^2-4\pi G\ro\;,
\end{align}
and, of course, $0\leqslant\mu\leqslant1$. We stress that $\omega_0^2$ is \emph{not} positive definite.

\section{Evolution of the density contrast and conditions for collapse}
The form \reff{master-system} of the evolution equations has the advantage that it clearly expresses the relationship between the physical quantities involved, other that being very well suited for numerical integration. Some further analytical insight can however be gained by reducing it to an unique higher-order equation for the variable $\delta(t)$. 

Considering the case of a matter-dominated Universe, and using the explicit time dependence of the quantities involved in the model in that case, \ie \erefs{time-parameters}, with some algebra one can derive the following fourth-order differential equation for $\delta(t)$:
\begin{align}\label{master-delta}
&9 t^4 \delta^{(4)}+
60 t^3 \delta^{(3)}+
\left[9\ld+76+9\lu t^{-2\nu}\right]\,t^2\,\delta^{(2)}+\nonumber\\
&\quad+\left[(12\ld+8)+12\lu(1-3\nu)\,t^{-2\nu}\right]\,t\,\delta^{(1)}+\\
&\qquad+\left[-6\ld\mu^2+3\lu(3\ld\mu^2+12\nu^2-2\nu)\,t^{-2\nu}\right]\,\delta=0\;,\nonumber
\end{align}
where $\delta^{(\ell)}$ denotes the $\ell^{th}$ derivative of $\delta$ with respect to time, and we have defined the following constants:
\begin{align}\label{Lambdas}
\nu\equiv \gamma-4/3\;,\quad
\lu=v_s^{2}k^{2}t^{2\gamma-2/3}\;,\quad
\ld=\omega_A^{2}t^{2}\;.
\end{align}
We recall that $\gamma$ is the specific-heat ratio ($P\sim\rho^{\gamma}$) and that $\gamma\geqslant4/3$, \ie $\nu\geqslant0$.

The most general solution of \eref{master-delta} for $\delta$ is found to be the superposition of four independent solutions $\delta_i$ $(i=1,\dots,4)$, given by:
\begin{align}\label{delta_sol}
\delta_i= A_i\;t^{x_i}\;\;
_{2}\mathcal{F}_{3}\Big[(a_{1i},\,a_{2i});(\,b_{1i},\,b_{2i},\,b_{3i});
\;-\frac{\;\lu t^{-2\nu}\,}{4\nu^{2}}\;\Big]\;,
\end{align}
where $_{p}\mathcal{F}_{q}[(a_{1},...,a_{p});(\,b_{1},...,b_{q});\,z]$ denotes the generalized hypergeometric function of argument $z$, the $A_i$'s are arbitrary integration constants and
\begin{subequations}
\begin{align}
&x_1=(-1+\sqrt{\Delta_-})/6\;,\qquad	x_2=(-1-\sqrt{\Delta_-})/6\;,\\
&x_3=(-1+\sqrt{\Delta_+})/6\;,\qquad	x_4=(-1-\sqrt{\Delta_+})/6\;,\\
&\phantom{\int_0^1}\!\!\!\!\!\Delta_\pm=13-18\ld\pm6\sqrt{(3\ld-2)^{2}+24\mu^{2}\ld\;}\;.
\end{align}
\end{subequations}
The constant coefficients $a$ and $b$ depend, in general, on $\nu$, $\ld$, $\mu$ and we report their complete expressions in \ref{Hyp_ab}.

We are now interested in discussing the asymptotic behavior of the hypergeometric functions in the limit of very small or very large argument, \ie $\lu/4\nu^{2}t^{2\nu}$$\gg$$1$ or $\ll$$1$. As in the non-magnetic case discussed in Ref. \cite{G&C}, we restrict the analysis to the range $0\leqslant\nu\leqslant1/3$, \ie treating the standard regime $4/3\leqslant\gamma\leqslant5/3$. 

From the asymptotic expansion of the $\mathcal{F}$ functions in the case of large argument, \ie $\lu/4\nu^{2}t^{2\nu}\gg1$, the density contrast always shows a damped oscillating behavior with time. In fact, in this regime
it always exists at least one asymptotic solution proportional
to positive power of the argument $\lu/4\nu^{2}t^{2\nu}$, which results to be the leading term of the solution superposition. In this case, $\delta$ decreases with time.

On the other hand, in the limit $\lu/4\nu^{2}t^{2\nu}\to0$, the asymptotic expansion of the solutions \reff{delta_sol},
can be written as
\begin{align}
\delta_i \sim  t^{x_i}+\mathcal{O}(\lu/4\nu^{2}t^{2\nu})\;.
\end{align}
In order for the gravitational collapse to occur, at least one of the modes has to be growing, \ie $x_i>0$.  It is fairly easy to show that $x_1$, $x_2$ and $x_4$ are always negative, whereas the sign of $x_3$ depends on $\mu$ and $\ld$. In particular, when $\mu\neq 0$, we obtain $x_3>0$ irregardless of the value of $\ld$, while when $\mu=0$, $x_3$ is positive if $\ld < 2/3$. 
This means that, on the plane orthogonal to the magnetic field ($\mu=0$), a new stability condition arise if the magnetic field is strong enough.

The threshold value related to $\Lambda_1$ that should discriminates the two regimes of growing and decreasing density contrast can be set as $\Lambda_1=1$ \cite{G&C}. Remembering that $\rho_0 = 1/6\pi G t^2$, such condition rewrites in terms of the wave number as
\begin{equation}
k \gtrless k_J \equiv \sqrt{\frac{24\pi G\, \nu^{2}\, \ro}{v_s^{2}}}\;,
\label{eq:Jeans}
\end{equation}
which is substantially the same as the usual Jeans condition for gravitational instability. In fact, in the non-magnetic case, (to which our analysis reduces for $\omega_A=0$) this is the only criterion that separates the growing and the decaying modes \cite{G&C}. 

In a similar way, the new threshold $\ld=2/3$ yields to the following condition
\begin{equation}
k \gtrless k_A \equiv \sqrt{\frac{4\pi G\,\ro}{v_A^{2}}}= \sqrt{\frac{16\pi^2 G \,\ro^2}{B_0^{2}}}\;.
\label{eq:JeansM}
\end{equation}

Summarizing, we find that the presence of a background magnetic field introduces an anisotropy in the stability criterion. While outside the plane orthogonal to $\bo$, the stability of the perturbations is dictated only by the standard Jeans condition $k\gtrless k_J$, on that plane the unstable modes are those for which the conditions $k < k_J$ and $k<k_A$ both hold\footnote{It is easy to verify how the physical meaning of the condition is that the timescale for gravitational collapse $\tau_c \sim L/v_c \sim \sqrt{L^3/GM}\sim\sqrt{1/G\ro}$ is much shorter than both the acoustic and Alfv\'en timescales $\tau_s \sim L/v_s \sim 1/k v_s$ and $\tau_A \sim L/v_A \sim 1/k v_A$.}. In other words, if $k_A<k_J$ (basically equivalent to $v_A > v_s$), there are Jeans-unstable modes (those in the window $k_A < k <k_J$) that, in the orthogonal plane, are stabilized by the magnetic pressure. The window of stable modes gets wider for larger values of the ambient magnetic field, as expected. We underline that these results are qualitatively the same as those obtained for a static and uniform background \cite{PCLMB11}. A similar analysis was carried on by the authors of Ref. \cite{VTP05} obtaining a similar results. However, their derivation contained a mistake when separating the real and imaginary components of the evolution equations \cite{G&C}. For this reason they find a second-order differential equation instead than the fourth-order one discussed here.

\section{Numerical Analysis}
In the previous Section, we have gained an important insight on the effect of a background magnetic field on the evolution of density perturbations. We now show some results obtained through the direct numerical integration of the differential system \reff{master-system}.
\begin{figure}
\centering
\includegraphics[width=0.75\hsize]{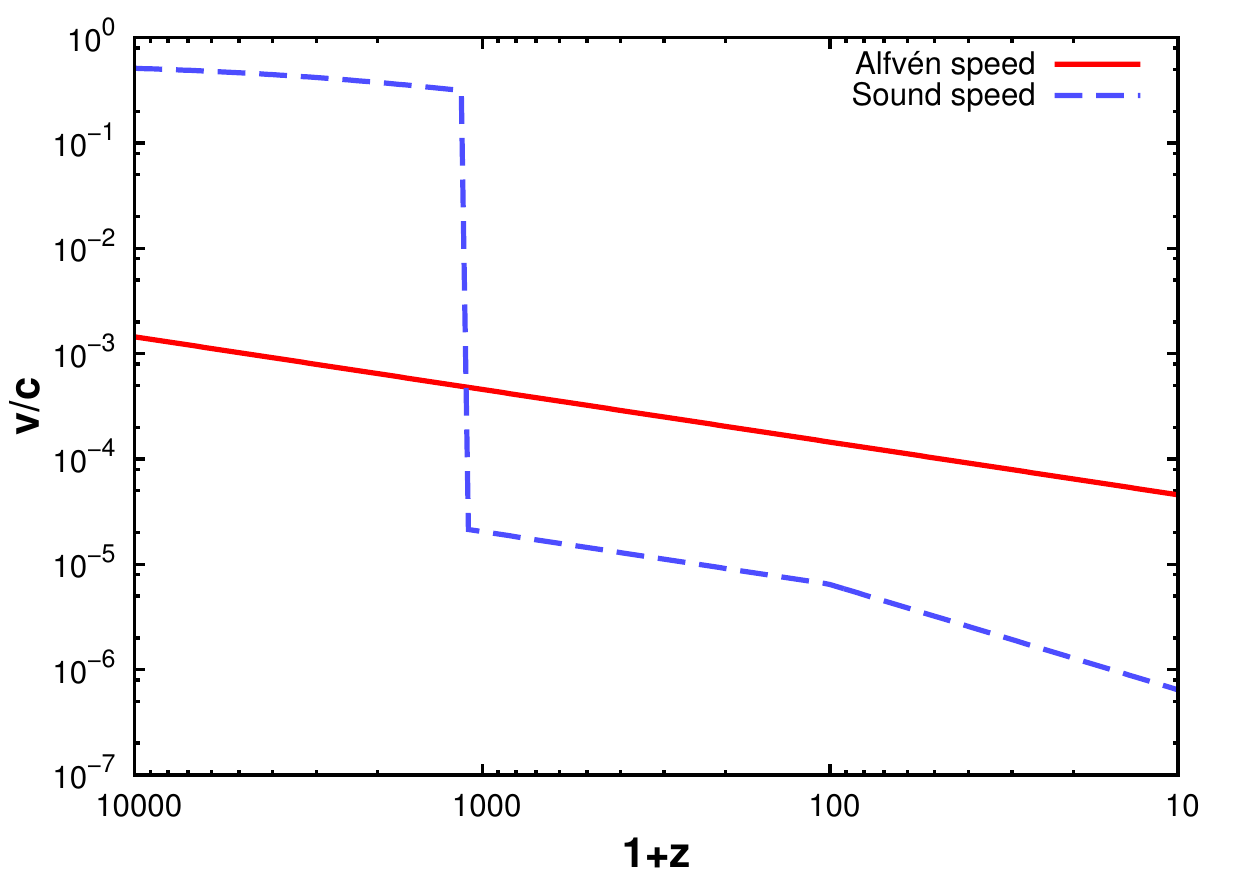}
\includegraphics[width=0.75\hsize]{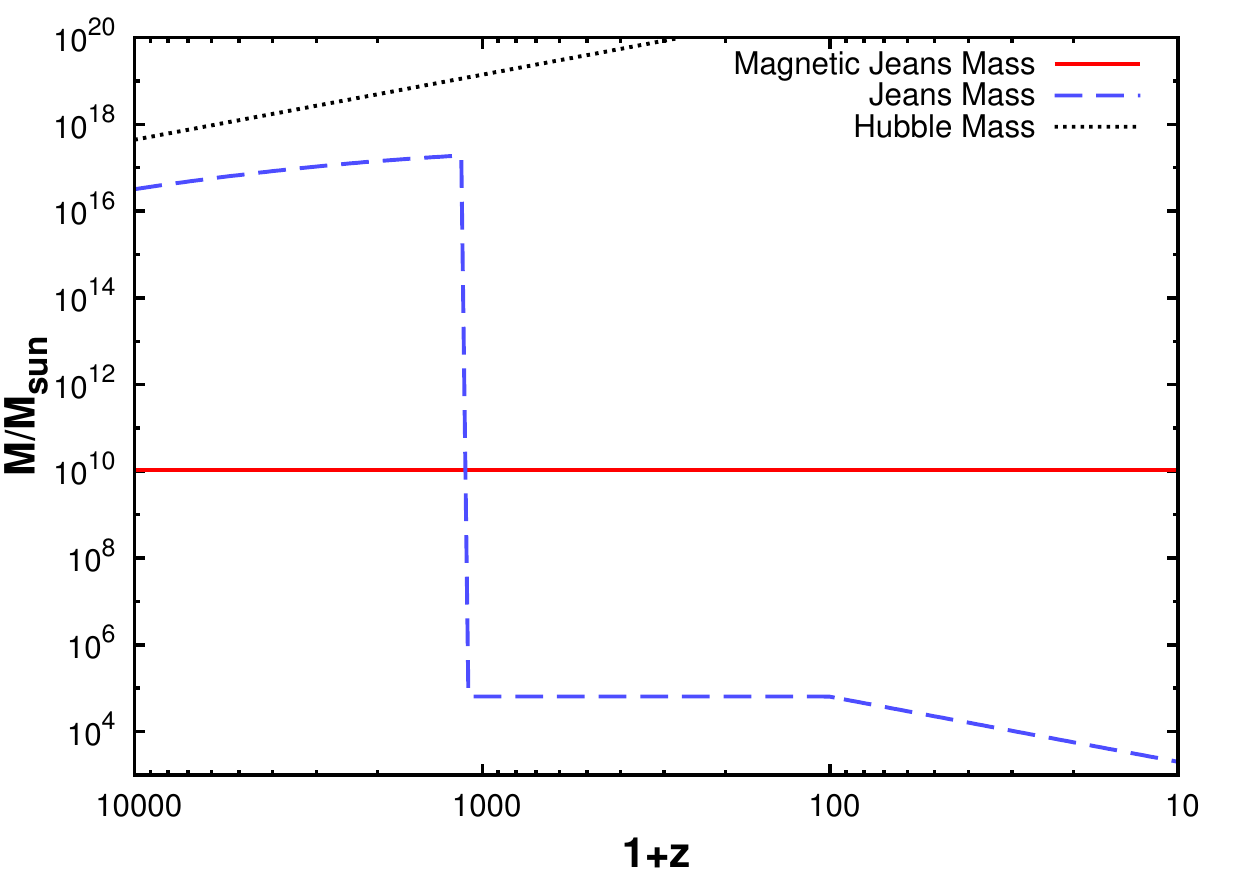}
\caption{Top panel: Alfv\'en (red solid line) and sound (blue dashed line) speed as functions of redshift, for $B_0(z=0)=10^{-9}$ G. The discontinuity at $z=1100$ corresponds to the recombination on neutral hydrogen. Bottom panel: magnetic (red solid line) and standard (blue dashed line) Jeans mass as a function of redshift. The mass contained inside the Hubble radius (black dotted line) is also shown for comparison. \label{fig:soundspeed}}
\end{figure}

\subsection{Preliminaries}
We will focus on the period of the cosmological evolution that goes from the onset of matter domination ($z\simeq 3000$) to the time of reionization ($z\simeq 10$). We start from matter domination because, before that time, the growth of density perturbation was slowed down and practically frozen by the rapid expansion of the Universe. In the matter-dominated Universe, $a\propto t^{2/3}$ and $H=2/3t$. We can ignore the presence of a dark energy component since this is sub-dominant until very recent times. The time period that we consider can be divided into two distinct phases, \ie before and after the recombination of hydrogen occurring at $\zrec=1100$. Before recombination, the baryons are completely ionized and they are tightly coupled to photons, at least at scales larger than the comoving photon mean free path $\lambda_\gamma\simeq 1.8 [(1+z)/(1+\zrec)]^{-2}$. At these scales, the total pressure of the fluid is given by radiation pressure. After recombination, most of the protons and electrons are in the form of neutral hydrogen atoms, leaving a small ionized fraction $x_e\simeq 2.5\times10^{-4}$. At the scales of interest, the neutral and ionized components are tightly coupled by collisions (see Sec. \ref{ssec:post_uni}) and can be treated as a single fluid. However, photons are now free streaming so that the baryon pressure is given just by kinetic pressure, dropping down by several orders of magnitude with respect to its pre-recombination value.
\begin{figure*}[ht]
\centering
\includegraphics[width=0.33\hsize]{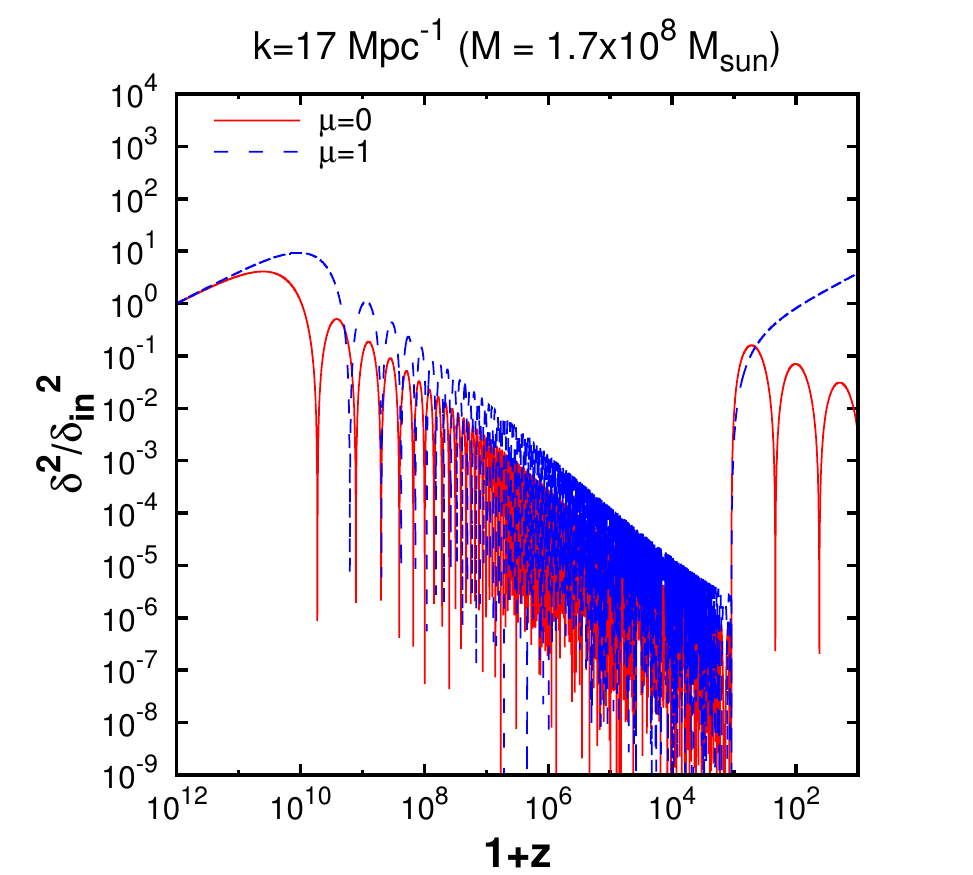}
\includegraphics[width=0.33\hsize]{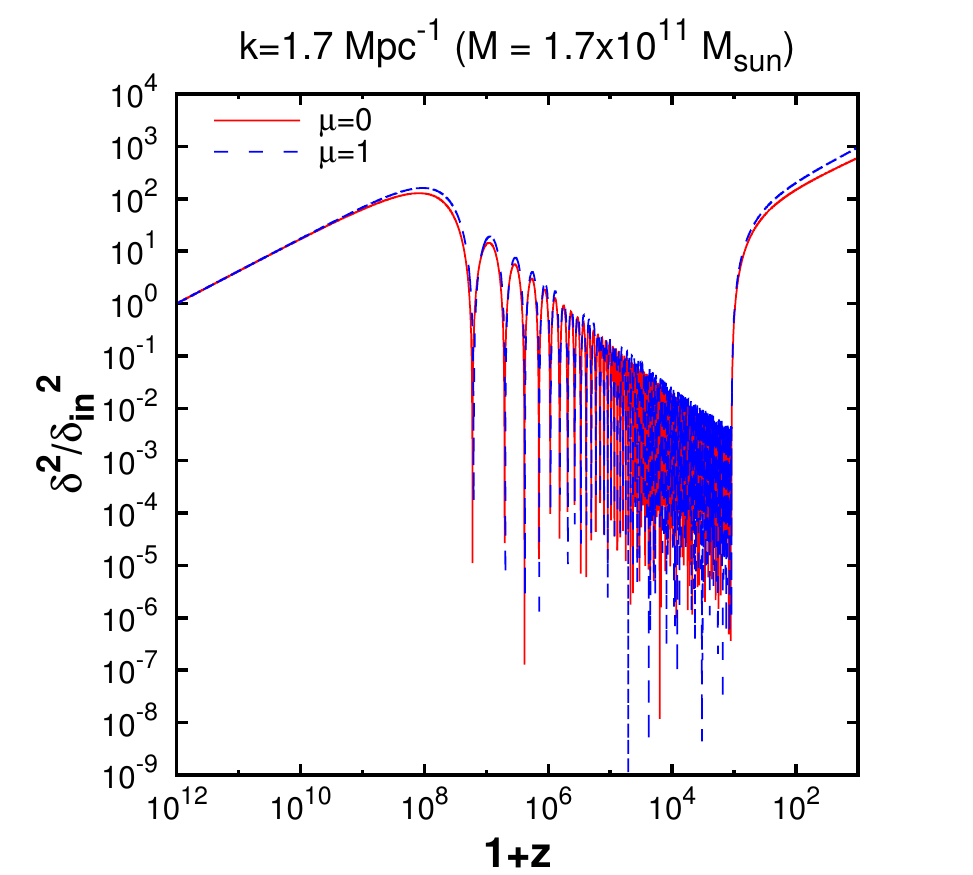}
\includegraphics[width=0.33\hsize]{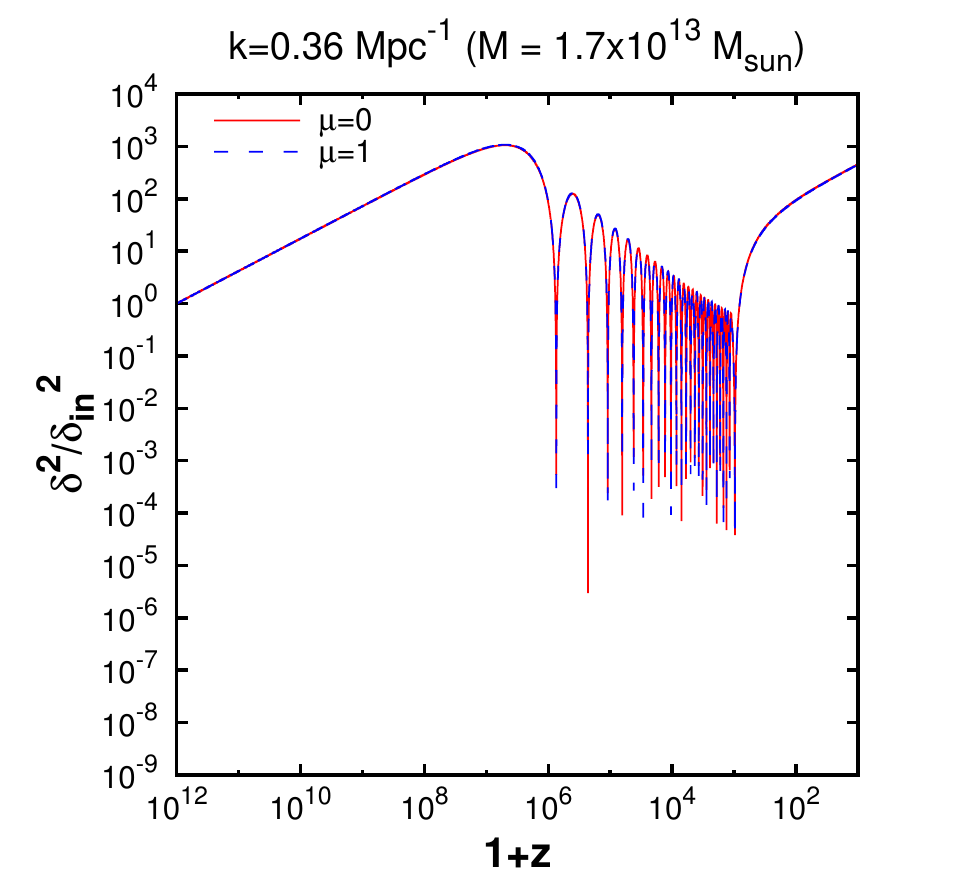}
\caption{Evolution of the dimensionless density perturbation $\delta$ with redshift, for $B_0(z=0)=10^{-9}$ G. The three panels show three different mass scales roughly corresponding, from left to right, to the dwarf galaxy, galaxy, and cluster scales. In each panel, we show the evolution of perturbations orthogonal (red solid line) and parallel (blue dashed line) to the background magnetic field.}
\label{fig:deltak}
\end{figure*}

In view of this, we take the speed of sound of the cosmological fluid before and after recombination to be \cite{G&C}
\begin{equation}\label{vsvs}
v_s^2|_{z>\zrec}=\frac{1}{3}\;\frac{k_B T_b \sigma}{m_p + k_B T_b \sigma}\;,\qquad
v_s^2|_{z<\zrec}=\frac{5}{3}\;\frac{k_B T_b}{m_p}\;,
\end{equation}
respectively, where  $\sigma = 4 a_{SB} T^3/3n_b k_B\simeq 1.5\times 10^9$ is the specific entropy. We recall that $T_b=T_\gamma=T_\gamma|_{z=0}(1+z)$ for $z>100$, while afterwards $T_b\propto (1+z)^2$. The expression above for the sound speed is rigorously valid only for scales corresponding to baryonic masses $\gtrsim 10^{11} M_\odot$, that stay above the photon mean free path until the time of recombination. At smaller scales, baryons lose the radiation support before recombination, roughly when the given scale goes below the photon mean free path, and it is at this time that the switch between the two expressions in \eref{vsvs} should take place. In the following, we shall consider masses nearly as small as $10^8\,M_\odot$, for which the photon decoupling effectively takes place at $z\simeq 3000$ (very close to the time of matter-radiation equality). However, we choose to switch between the two expressions for the sound speed at $z=\zrec$ irregardless of scale and shall comment later on how we expect this choice to affect our results.

Following the discussion in Sec. \ref{ssec:post_uni}, the Alfv\'en velocity is taken to be
\begin{equation}
v_A=\sqrt{\frac{B_0^2}{4\pi\rho_b}}\;,
\end{equation}
where $\rho_b$ should always be intended as the total baryon density, both before and after recombination.

A plot of the Alfv\'en and sound speeds as functions of redshift is shown in the upper panel of Figure \ref{fig:soundspeed}. The sudden drop in the sound speed at recombination is due to the sharp decrease of baryon pressure after photon decoupling.

In a detailed model, the presence of different uncoupled components making up the matter content of the Universe should be taken into account. In fact, most of the matter ($\sim 80\%$) is in the form of cold dark matter (CDM), interacting with the baryon-photon fluid only through the gravitational force. Thus, a proper treatment should rely on a two-fluid description. In the following, we shall ignore perturbations in the CDM component, however we argue that we can still draw meaningful conclusions about the perturbations in the baryonic component. In fact, in the pre-recombination era, the large radiation pressure prevents baryons to fall into the potential wells created by CDM; this is known to be true in the non-magnetic case but we expect it to hold also in the case under consideration since, as seen in the last Section, the magnetic field only acts to increase stability. Thus in this regime the CDM and baryon perturbations are effectively decoupled. After recombination, the baryon density perturbations will take some time to catch up with those in the CDM component and we expect our treatment will rigorously remain valid for some time.

Before showing the results of the numerical integration, we illustrate in the lower panel Figure \ref{fig:soundspeed} the evolution of the standard Jeans wavenumber \reff{eq:Jeans} and of its ``magnetic'' counterpart \reff{eq:JeansM}. In order to take out the change in $k_J$ and $k_A$ due to the expansion, we follow the convention to express the results in terms of the mass contained inside the corresponding length scales $1/k_{J,A}$. In particular, we consider the total baryonic mass (irrespective of the ionization state), contained inside a sphere of radius $2\pi /k_{J,A}$. It can be seen that the window of modes that are made stable by the magnetic field, \ie those between the red dashed and the black solid line, spans, right after recombination, five orders of magnitude in mass.

As noted in the introduction, the existence of a magnetic Jeans length has been studied previously and all expressions for the critical wavenumber agree, apart from numerical factors, with expression \reff{eq:JeansM}. However, the numerical estimates of this and associated quantities that are found in the literature sometimes differ from our results. The reason seems to be that often the density $\rho_0$ that appears in \eref{eq:JeansM} is taken to be the present critical density $\rho_{c}\simeq 9\times 10^{-27}$ kg m$^{-3}$. This yields at the present time a magnetic Jeans length $\lambda_A \sim 1/k_A \sim 10$ kpc for $B_0(z=0)=10^{-9}$G and $\lambda_A\sim 1$ Mpc for $10^{-7}$G \footnote{The latter value has also been said to be of the order of the scale of a galaxy cluster, while in effect it is closer to the scale of a galaxy, as it can be seen by the fact that the mass enclosed inside a sphere of 1 Mpc radius at the critical density is  $\sim 10^{11} M_\odot$. The reason why a galaxy is much smaller than 1 Mpc is that it has detached from the Hubble flow and undergone non-linear evolution, so that its density is much larger than the cosmological average \cite{E&U}.}. Using instead the baryon density for $\rho_0$ will yield values of $\lambda_A$ a factor $\rho_c/\rho_b=\Omega_b^{-1}\simeq 20$ larger, \emph{i.e.}, $\lambda_A\sim 0.2\,(20)$ Mpc for $B_0(z=0)=10^{-9}\,(10^{-7})$ G. This amounts to a factor $20^3\simeq 10^4$ difference in the corresponding mass scale\footnote{The discussion so far has ignored the gravitational action of dark matter; this can be roughly taken into account by using the total matter density $\rho_m$ at the numerator of \eref{eq:JeansM}, while keeping the same expression for $v_A$. This makes the value of $\lambda_A$ roughly twice smaller than in the case in which only baryons are considered, \emph{i.e.}, $\lambda_A\sim 0.1\,(10)$ Mpc for $B_0(z=0)=10^{-9}\,(10^{-7})$ G.}.

\subsection{Results}
\begin{figure*}[ht!]
\centering
\includegraphics[width=\hsize]{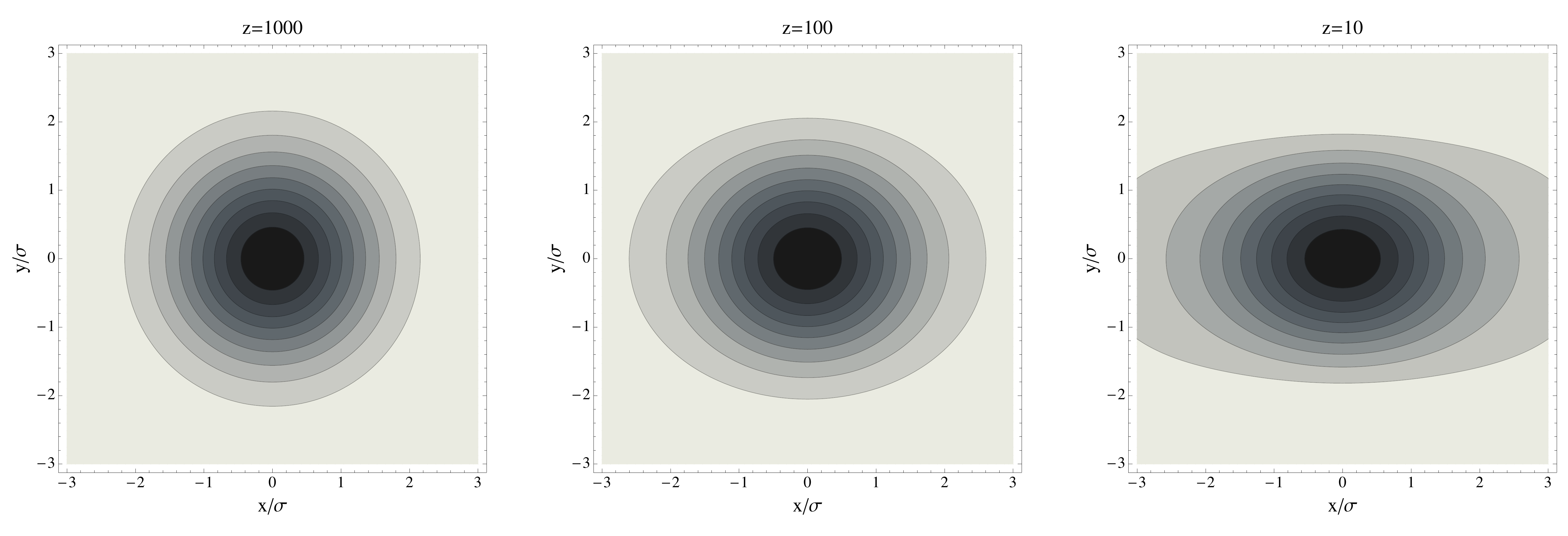}
\caption{Equal density contours of an initially spherically symmetric Gaussian perturbation at the scale of a dwarf galaxy in the $x$-$y$ plane, at different times. The contours correspond to $(0.1,\,0.2,\,\dots,\,0.9)$ times the central density. The background magnetic field is directed along the $y$ axis and $B_0(z=0)=10^{-9}$ G.}
\label{fig:deltax}
\end{figure*}
We now discuss the results of the direct numerical integration of \erefs{master-system}. The initial conditions for the integration have been chosen using the fact that power-law solutions for $\delta$ can be found in the limit $t\to 0$. There are four distinct solutions of this kind, but only one corresponds to a growing mode. We have matched the initial conditions to the asymptotic growing solution at the initial time of integration. The latter has been chosen so that all the modes of interest were outside the horizon at that time. Even if the initial time falls in what would be the radiation-dominated era, nevertheless we always consider a matter-dominated Universe. All results have been normalized to the initial value of the density contrast.
\begin{figure*}
\centering
\includegraphics[width=0.8\hsize]{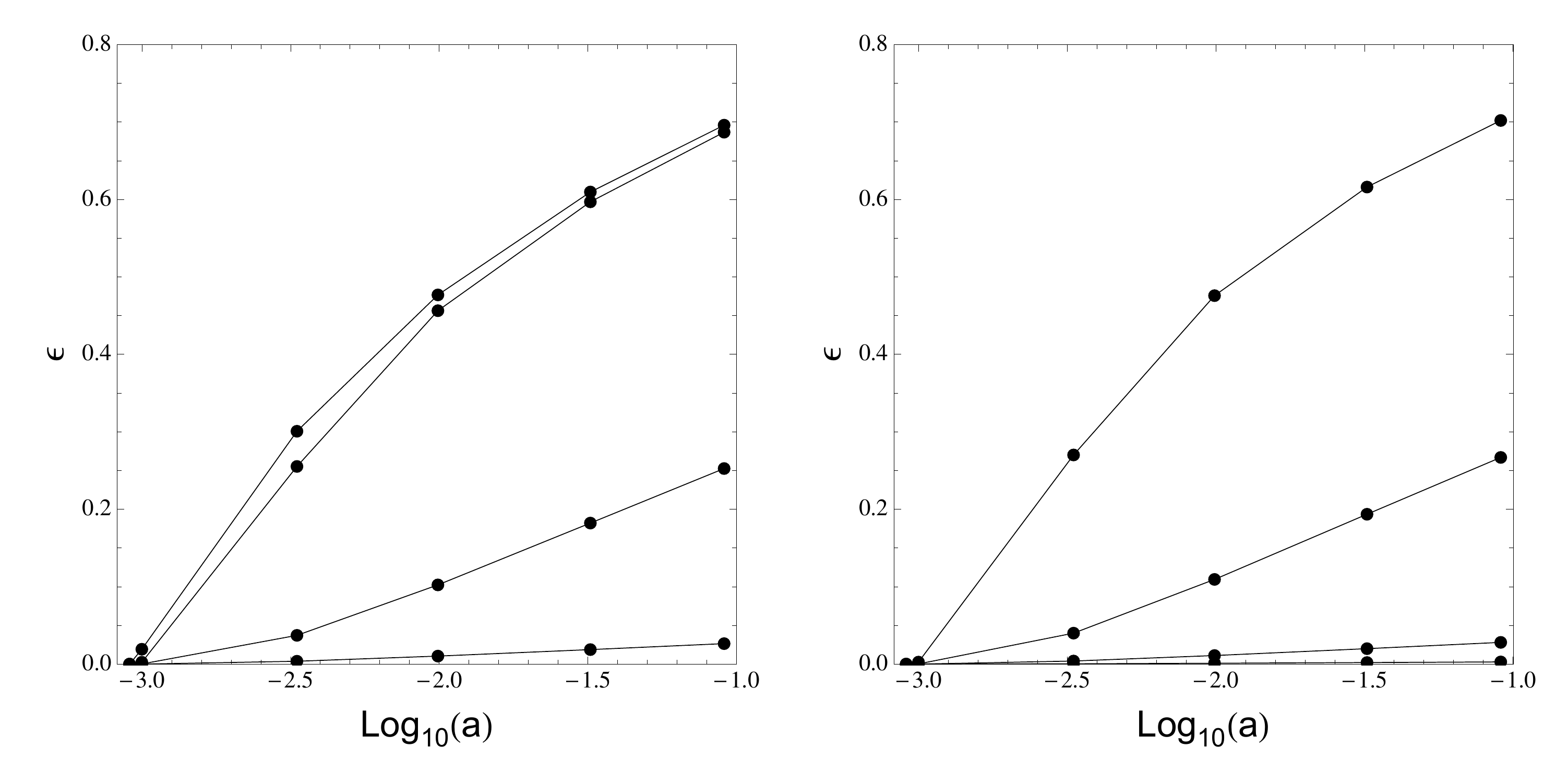}
\caption{Left panel: Eccentricity $\epsilon$ of a Gaussian perturbation at the scale of a dwarf galaxy as a function of redshift. The background magnetic field increases by a factor 10 with each curve, starting from $B_0(z=0)=10^{-11}$ G (bottom curve) up to $B_0=10^{-8}$ G (top curve). Right panel: the same as the left panel, but for a perturbation at the galactic scale.}
\label{fig:ecc}
\end{figure*}

In Figure \ref{fig:deltak}, we show the evolution of the density contrast for three different wavenumbers $k=(17,\,1.7,\,0.36)\,\mathrm{Mpc}^{-1}$ (normalized at the present time), \ie for the following baryonic masses $M=(1.7\times 10^{8},\,1.7\times 10^{11},\,1.7\times 10^{13})\, M_\odot$. These masses roughly correspond to the scale of a dwarf galaxy, of a galaxy and of a galaxy cluster respectively. For each mode, we show the evolution in both the direction parallel to the background magnetic field ($\mu = 1$) and orthogonally to that direction ($\mu=0$). In all cases, the perturbation is initially growing but then starts to oscillate once the Jeans mass (that is growing) becomes larger than the mass of the perturbation. This happens earlier for smaller scales. In this phase the magnetic pressure does not play any role, as the much larger radiation pressure is actually providing the force that prevents the collapse. In fact, there is no difference in the evolution parallel and orthogonal to the field, as the radiation pressure is isotropic. The situation changes dramatically after recombination, when the baryon pressure drops and only the magnetic pressure can possibly oppose the growth, at least in the plane orthogonal to the field. Thus, perturbations in the direction of the field can grow unhindered, while the perturbations that are orthogonal can be stabilized. As it can be seen from Figure \ref{fig:deltak}, this is what happens for perturbations at the dwarf galaxy scale: at $z=10$, the relative growth of parallel perturbations with respect to orthogonal ones is of order 100. For perturbations at the galactic scale and larger, instead, the evolution is basically the same in all directions. This can be understood by looking at the lower panel of Figure \reff{fig:soundspeed}, from which it is clear that the pressure induced by a magnetic field of $10^{-9}$ G can only stabilize perturbations with mass $\lesssim 10^{10}\, M_\odot$.

We recall that we have neglected the fact that, at the dwarf galaxy scale, the support of radiation pressure is not lost at recombination but some time before (see previous Subsection). From the discussion above, it is clear that, had we taken into account this fact, the evolution of orthogonal and parallel perturbations would have begun to differentiate earlier. This goes in the direction of enhancing the anisotropic growth of perturbation and the ``squeezing'' effect studied in the following.

The fact that, after recombination, the evolution of the density contrast in the presence of a magnetic field changes for different directions leads to the reasonable expectation that some degree of anisotropy will be generated even in initially symmetric structures. In order to show this, we consider a Gaussian density fluctuation with standard deviation $\sigma$ in coordinate space at recombination:
\begin{equation}
\delta(\vec x,\,t_{rec})= \delta(\vec x=0,\,t_{rec}) e^{-\frac{|\vec x|^2}{2\sigma^2}}\;,
\end{equation}
where the $\vec x$ are comoving coordinates centered at the maximum of the perturbation. After Fourier-transforming, we separately evolve the different harmonics in momentum space using \erefs{master-system} and we finally transform back to obtain the perturbation in coordinate space at a later time. In Figure \ref{fig:deltax} we show, for a background magnetic field directed along the $y$ axis and with a present intensity of 1 nG, the evolution of a perturbation with $\sigma = 0.05$ kpc at recombination (so that the $3\sigma$ region encloses a mass $M\simeq1.5\times 10^{8} M_\odot$ at the mean baryonic density, \ie roughly the mass of a dwarf galaxy). In particular, we show equal density contours at $z=1000,\,100$ and $10$. It is evident from the figure how the perturbation becomes progressively squeezed along the direction orthogonal to the magnetic field.

In order to quantify the anisotropy in the perturbation, we consider the isodensity contour corresponding to half the value at the peak and calculate its eccentricity $ \epsilon = \sqrt{1-b^2/a^2}$, where $a$ and $b$ are the lengths of the semi-major and semi-minor axis of the contour, respectively. In Figure \ref{fig:ecc}, we show how the eccentricity changes with redshift; for the parameters used above, we get $\epsilon \simeq 0.7$ at $z=10$.

\section{Conclusions}

We have studied the effect of a background magnetic field on the linear evolution of cosmological density perturbations at scales well below the Hubble length, where a Newtonian treatment can be used, focusing on the matter-dominated era. The conditions that allow for the growth of small density perturbations have been clearly stated. In particular, we have found that in the plane orthogonal to the ambient magnetic field, a new critical length appears, related to the presence of the magnetic pressure, while everywhere else outside that plane the stability is dictated by the standard Jeans criterion. This is also confirmed through a direct numerical integration of the relevant MHD equations during the matter-dominated era, and this effect is shown to be possibly important after recombination, when the magnetic pressure of baryons is much larger than their the kinetic pressure. Finally, it has been shown how the dependence of the critical scale on the angle between the perturbation wavevector and the magnetic field could lead to a sizable anisotropy in the perturbations at sub-galactic scales at the onset of non-linearity.

Our analysis has relied on some approximations: in particular, we have ignored the gravitational effects of dark matter perturbations. We have argued that this approximation limit the validity of our treatment to some time after recombination. We defer a more detailed and fully general relativistic analysis, also taking into account the different fluid components, to a future work.

{\small \textbf{Acknowledgment}: N.C. gratefully acknowledges the \emph{CPT - Universit\'e de la Mediterran\'ee Aix-Marseille 2} and the financial support from ``Sapienza'' University of Roma. M.L. acknowledges  support from a joint \emph{Accademia dei Lincei / Royal Society} fellowship for Astronomy. The work of M.L. has been supported by \emph{MIUR - Ministero dell'Istruzione, dell'Universit\`a e della Ricerca} through the PRIN grants ``Matter-antimatter asymmetry, Dark Matter and Dark Energy in the LHC era'' (contract number PRIN 2008NR3EBK-005) and "Galactic and extragalactic polarized microwave emission" (contract number PRIN 2009XZ54H2-002).}

\newpage
\appendix
\section{Hypergeometric Coefficients}\label{Hyp_ab}
In following, we write the complete form of the coefficients of the hypergeometric function of \eref{delta_sol}. They read: 
\begin{subequations}
\begin{align}
a_{1(^{1}\;_{2})}&=1\mp\sqrt{\Delta_{-}\;}/12\nu-\sqrt{1-36\mu^{2}\ld\;}/12\nu\;,\\
a_{1(^{3}\;_{4})}&=1\mp\sqrt{\Delta_{+}\;}/12\nu-\sqrt{1-36\mu^{2}\ld\;}/12\nu\;,\\
a_{2(^{1}\;_{2})}&=1\mp\sqrt{\Delta_{-}\;}/12\nu+\sqrt{1-36\mu^{2}\ld\;}/12\nu\;,\\
a_{2(^{3}\;_{4})}&=1\mp\sqrt{\Delta_{+}\;}/12\nu+\sqrt{1-36\mu^{2}\ld\;}/12\nu\;,
\end{align}
\end{subequations}
and
\begin{subequations}
\begin{align}
b_{1(^{1}\;_{2})}&=1\mp\sqrt{\Delta_{-}\;}/6\nu\;,\\
b_{2(^{1}\;_{2})}&=1\mp\sqrt{\Delta_{-}\;}/12\nu-\sqrt{\Delta_{+}\;}/12\nu\;,\\
b_{3(^{1}\;_{2})}&=1\mp\sqrt{\Delta_{-}\;}/12\nu+\sqrt{\Delta_{+}\;}/12\nu\;,\\
b_{1(\;_{3})}&=1-\sqrt{\Delta_{+}\;}/6\nu\;,\\
b_{2(\;_{3})}&=1-\sqrt{\Delta_{-}\;}/12\nu-\sqrt{\Delta_{+}\;}/12\nu\;,\\
b_{3(\;_{3})}&=1+\sqrt{\Delta_{-}\;}/12\nu-\sqrt{\Delta_{+}\;}/12\nu\;,\\
b_{1(\;_{4})}&=1-\sqrt{\Delta_{-}\;}/12\nu+\sqrt{\Delta_{+}\;}/12\nu\;,\\
b_{2(\;_{4})}&=1+\sqrt{\Delta_{-}\;}/12\nu+\sqrt{\Delta_{+}\;}/12\nu\;,\\
b_{3(\;_{4})}&=1+\sqrt{\Delta_{+}\;}/6\nu\;.
\end{align}
\end{subequations}

\end{document}